# DevOps Automation Pipeline Deployment with IaC (Infrastructure as Code)

Adarsh Saxena
*Department of Electronics and Communication*
*University of Allahabad*
Prayagraj, India
adarshsaxena358@gmail.com

Sudhakar Singh
*Department of Electronics and Communication*
*University of Allahabad*
Prayagraj, India
sudhakar@allduniv.ac.in

Shiv Prakash
*Department of Electronics and Communication*
*University of Allahabad*
Prayagraj, India
shivprakash@allduniv.ac.in

Tiansheng Yang
*Faculty of Business and Creative Industries*
*University of South Wales*
Pontypridd, United Kingdom
tiansheng.yang1@southwales.ac.uk

Rajkumar Singh Rathore
*Cardiff School of Technologies*
*Cardiff Metropolitan University*
Cardiff, United Kingdom
rsrathore@cardiffmet.ac.uk

*Abstract*—DevOps pipeline is a set of automated tasks or processes or jobs that has tasks assigned to execute automatically that allow the Development team and Operations team to collaborate for building and deployment of the software or services. DevOps as a culture includes better collaboration between different teams within an organization and the removal of silos between them. This paper aims to streamline the current software development and deployment process that is being followed in most of today's generation DevOps deployment as Continuous Integration and Continuous Delivery (CI/CD) pipelines. Centered to the level of software development life cycle (SDLC), it also describes the current ambiguous definition to clarify the implementation of DevOps in practice along a sample CI/CD pipeline deployment. The further objective of the paper is to demonstrate the implementation strategy of DevOps Infrastructure as Code (IaC) and Pipeline as a code and the removal of ambiguity in the definition of DevOps Infrastructure as a Code methodology.

## I. Introduction

Often in many organizations, due to ambiguity and conflict in the DevOps [1] description, it is always unclear how to implement the concept in practice. Thus, the implementation process in DevOps is much more crucial compared to the concept and definition itself. It is important to streamline the current software development and deployment process that is being followed in the most of present deployment pipelines. In the current scenario, most of the development and deployment of software or services require different teams and stairs of processes. Therefore, the creation of silos between different teams is obvious most of the time. The requirement of this work can be clearly identified to remove the silos between different teams of an organization using the DevOps strategy and pipeline model.

From the top level, DevOps seems to be a practice of operations and the Development engineers participating and contributing together in the entire lifecycle of the software or other IT product. DevOps has its values, principles already defined, and methods, practices, and tools to use for its implementation. But still, several times in many organizations, the ambiguity arises in the implementation of DevOps practices and strategies. Even after knowing about DevOps and its strategies, it is still difficult for most of organizations to find the best approach according to the project requirement.

This paper aims to solve the problem of silos between different processes and teams that are involved in the development or deployment of software or service. It resolves the ambiguity and conflict in the description of DevOps by demonstrating how the DevOps Pipeline works in practice with a sample pipeline creation. Further, it clarifies the implementation strategy of the DevOps pipeline for increasing the automation and faster release time of the software, with special emphasis on the Infrastructure as a Code and Pipeline as a Code.

This paper starts with an introduction to the DevOps Automation Pipeline Deployment. In the subsequent sections, various discussions are available along with the required information and examples. In Section 2, an analysis and discussion of previous works in this field is discussed. In Sections 3 and 4, detailed design and implementation of the work are described. In Section 5, the working and testing of the implementation is presented. Last but not least, Chapter 6 covers the conclusion and scope of the future works.

## II. Literature Review

In the current scenario of software engineering and development, DevOps is not a new term. DevOps can be described or defined as a culture and philosophy of the software engineering branch that exploits cross-functional teams for building, testing, and releasing the software products faster with more reliability through the automation [2]. This is a very generalized definition, and the DevOps definition changes in every other research paper/report and from one organization to another.

At first, DevOps definition is clear to understand and grasp but from the practitioner's point of view, the DevOps definition and description are not always enough to communicate its real meaning. This ambiguous situation arises because most of the time, DevOps description does not clarify from the implementation perspective. The real meaning of the DevOps comes only when it is implemented within an organization. Generally, the DevOps practices are interchangeably understood with that of end-to-end automation of pipeline for software development or creating workflow of the shortest possible time or sometimes just associated it as a culture to follow [3]. DevOps implementation has three main modes as follows [2], [4].

- DevOps as a culture practice
- DevOps as a responsibility or position
- DevOps as a collection of CI/CD Processes

DevOps as a culture includes better collaboration between different teams within an organization and removal of silos between them. Whereas the Continuous Integration and Continuous Delivery (CI/CD) practices include pipeline deployment, automation, etc. In the case of DevOps as a position, it creates a separate responsibility/position in terms of DevOps practitioners. The literature review [5] emphasizes in detail about the capabilities and areas of DevOps.

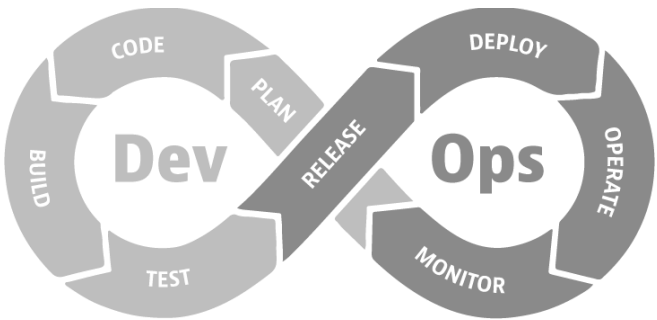

Fig. 1. DevOps Cycle [1] [6], [7].

DevOps strategy is not only levied on the technical part of an organization but on the management team as well. DevOps has a specific set of values and principles defined for an organization to be followed. Most of the time, we associate DevOps as more of a technology, and the word DevOps gets turned or reduced to technology is a manifestation of how desperately we need a cultural shift. The culture within an organization is part of the DevOps core values. The main aim of the DevOps, for which it was derived and created, was to remove the silos between the development team and the operations team. Thus, the culture within an organization performs a greater role in the implementation and enactment of the DevOps Core values. The culture is driven by the behavior of each and every person working for the organization and the mutual understanding between people of different teams like development and operation teams.

In the same way, the automation of the people, processes, and tools is also part of the DevOps core values. For example, the automation of tools can be done using IaC (Infrastructure as Code) or Pipeline as Code, which we are addressing as one of the parts of this paper. The core values of DevOps also include the measurement, which involves metrics and logs creation of different tools and processes. The metrics help engage teams and the overall goal of the organization [1].

Moreover, there are some principles in the DevOps that are defined like Systems Thinking, Amplifying Feedback loops, etc. Systems thinking tells us that the focus must be on the overall outcome and product of the entire pipeline or value chain. Amplifying feedback loops means short and effective feedback loops are the key aspects to prolific product development, software development, and the operations. For instance, a bug found at the development phase is better than the same bug found at the testing phase because in the former case, it will take less time for it to be resolved as illustrated in Fig. 2.

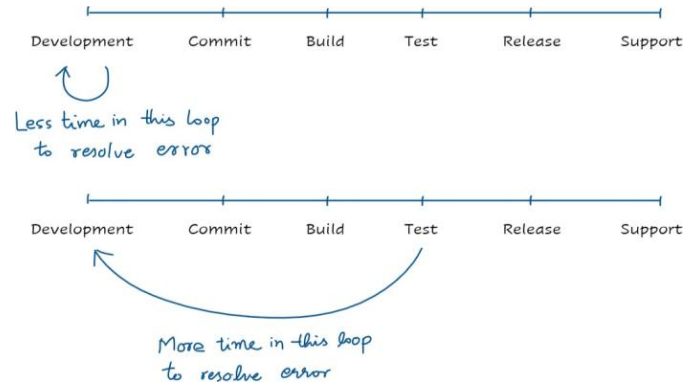

Fig. 2. Amplifying Feedback loops

When we go into the DevOps methodologies, the first one that comes up in the list is the Continuous Delivery, which is the practice of coding, testing, and releasing the software product frequently and in small batches to enhance the overall quality and velocity [8]. Another methodology is the infrastructure as a Code, which says that the systems can and should be treated or used as the code. It also comes up with several advantages for an organization, for example, the whole infrastructure can be replicated anywhere, anytime because all its configuration can be managed using code. In the same way, there are other methodologies defined as well like lean management, visible ops- style change control, etc.

After gathering the results from the number of surveys of DevOps practices in the market [9]–[11], the following points can be concluded.

- 49% agree that they see a faster time to market.
- 49% agree that it improved their deployment frequency.
- 61% agree that it helped them produce higher quality deliverables.
- 85% people face difficulties in implementation of DevOps practices in their organization.

It is very clear from the such surveys that the implementation of DevOps is most of the time messed up within the organization. The main reasons behind this are lack of skills in the employees or proper knowledge about its implementation, legacy infrastructure or adjusting to corporate culture vs DevOps culture, and orchestrating automation etc. [12].

III. METHODOLOGY AND IMPLEMENTATION

A CI/CD pipeline is generally a set of Jobs that can be parallel or sequential depending upon the use case of the organization or software. These jobs start to run one by one as soon as the pipeline is triggered. Generally, the CI/CD pipelines are always sequential, and many pipeline creation tools do not allow or use parallel computing or parallel Job implementation. We propose a parallel implementation which can be implemented using Scripts and can be executed parallelly within one job of the pipeline.

DevOps implementation in practice involves the deployment of CI/CD pipelines. CI/CD pipelines can be designed in a number of ways depending on the requirement and use case. In this paper, a pipeline is developed that automatically detects whenever the code is pushed to the central repository (or GitHub repository), pulls the code automatically, then does the automated testing in the development server and deploys the code to the production server. The pipeline is created using Jenkins (a CI/CD pipeline creation tool) [13], and Git and GitHub are used as source code management (SCM) tool for version controlling and code management [14]. Automation scripts are written using shell scripting. Most importantly, in Jenkins, the Pipeline as a Code can be used, which brings in the best out of DevOps. Django is used to create a sample web app for the demonstration of the whole process while httpd Server is used for the deployment of the sample site (for testing the working). AWS (Amazon Web Services) cloud is used for the deployment of the web app to the server, moreover, the Jenkins server is also running on the AWS Instance (a minimum of t2.micro instance is required) [15]. Python is used for writing scripts in between the pipeline. Terraform is used for the infrastructure creation using IaC (Infrastructure as Code) [16], [17].

Here, the approach for the deployment of the DevOps pipeline is based on the roots of the Infrastructure as a Code and the pipeline as a code strategy. For the demonstration of the DevOps implementation strategy using Infrastructure as a Code and Pipeline as a Code with Terraform (HCL) and Jenkins (Groovy) respectively, a sample project is developed. In the project, the first step is to set up the Jenkins server. For this, we used the IaC with Terraform which first creates an instance in the AWS and then set up the Jenkins inside of that instance. The code for the IaC is written in the HCL (Hashicorp Configuration Language) [16], [17]. While creating the instance using the HCL code, the web server can also be configured since we need the web server in the future. Then, the pipeline is created using the pipeline as a code. The pipeline created has the jobs to pull the code from the GitHub and then, deploy the same in the server.

The workflow of the implementation/operation of the CI/CD pipeline starts from the development of the software/service and goes towards deployment after testing and other elemental phases included in the deployment process. In the beginning, the team of developers develop the app or the software. At this phase, developers are free to follow any of the software development practices regarding the creation of software. They can use SCRUM and AGILE methodologies [18] as well during development as it just adds up in the DevOps implementation. After the development of the first version of the software, which is ready to be released, it is first pushed to the central repository of the organization where the code is pulled up by the pipeline automatically. As soon as the code is committed to the central repository, it triggers the pipeline to start implementing its Jobs one by one. In our example, as shown in Fig. 3, the code is being pulled by the first Job of the pipeline.

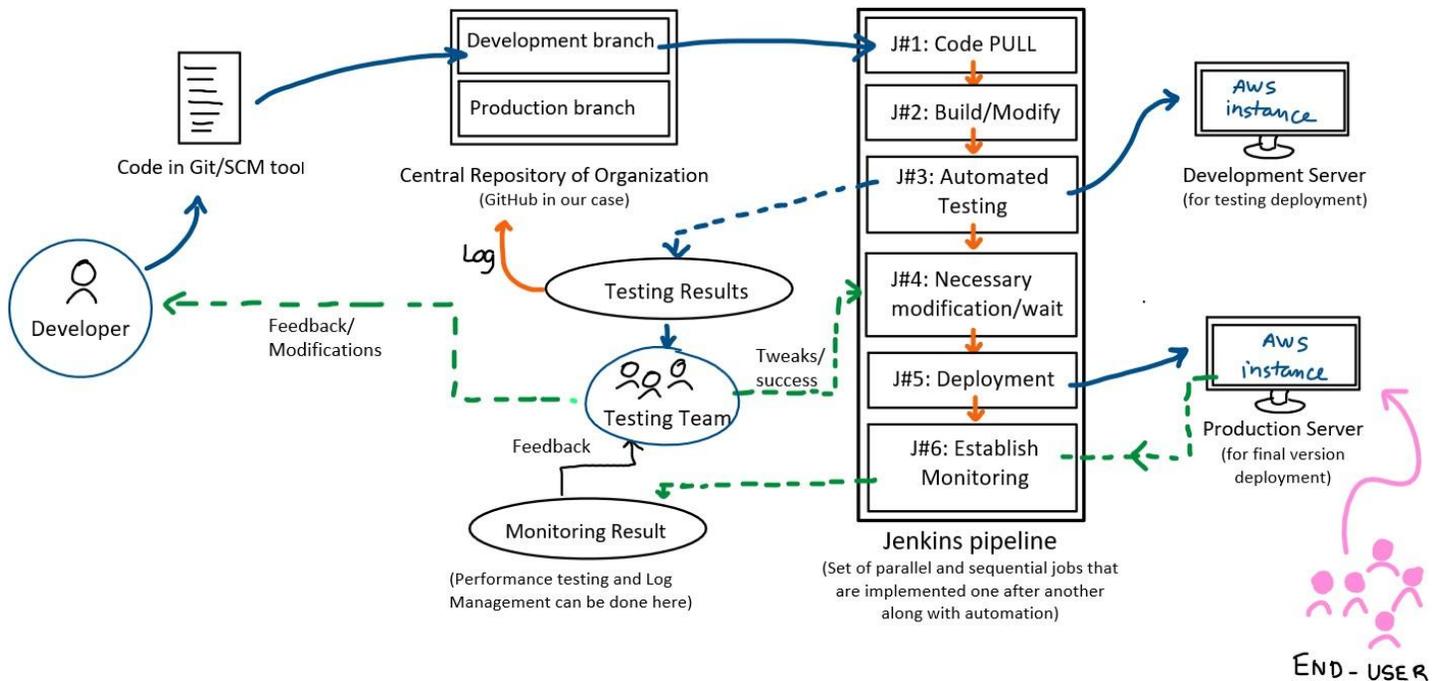

Fig. 3. Detailed design of the DevOps pipeline

After the code is pulled by Job J#1, it will trigger the next job whose role is to build or modify the code for deployment and testing. At Job J#3, automated testing on the software will start executing. This testing Job can also be integrated with manual testing or manual verification of automated testing results. Bringing in the manual authority is surely a decremental step towards automation, but it will provide better control and flexibility to the testing regarding the security of the software/service.

At the automated testing job, most of the time, the job needs to deploy the build software to the development server where it can perform testing. As the testing is dynamic in nature, i.e., it needs to deploy the service for that time only in which it is performing tests. Thus, the development server can take advantage of docker to create an instance and launch the service over there.

After the testing is completed, the results are then passed to the testing team (if manual operations are customized) otherwise, it will automatically take the necessary actions. On successful completion of testing jobs, the pipeline proceeds towards the deployment phase where it will finally deploy the whole code to the production server.

Before implementing anything in the pipeline, the administrator needs to create a pipeline and before creating a pipeline, the admin needs to set up the respective pipeline creation tools like Jenkins. The focus of the paper is to implement as much automation as possible at every phase, wherever possible, be it a development or the deployment phase. The automation can only be increased and implemented through code using which anything can be replicated any number of times. The first step in the implementation requires to set up the Jenkins server in an instance.

Setting up of Jenkins involves three steps. First is the installation of dependencies, which include the OpenJDK. The next step involves the installation of the Jenkins software. The last step is to start the services of the Jenkins server. Though these steps are quick and easy but they need to be implemented using Terraform, so that, if we need to recreate or replicate this in the future then we can do it with a single command. In Terraform, we first need to create the instance in the AWS along with the necessary configurations. Then, we need to set up the Jenkins as discussed earlier inside the instance created, using the HCL code only. The IP and the password of the Jenkins server which is to be entered at the starting can be seen as the output of the "terraform apply" command of Terraform.

After the successful deployment of the Jenkins server in the AWS instance, we can start creating the pipeline inside the Jenkins server using the pipeline as a code. Jenkins provides a pipeline as a code feature by using the Groovy language to write the pipeline code. The pipeline as a code implementation makes it easy to replicate the pipeline really fast if needed in the future.

IV. EXPERIMENTATION AND DISCUSSION

The pipeline demonstrated here can be implemented in a system with specifications higher than or equal to 1 core CPU and a minimum of 2 GiB of RAM in any Linux Distribution. Tasks that involve higher compute power for their implementation within the pipeline may result in a crash of the entire pipeline or respective pipeline creation tools used. The working/processing of the CI/CD pipeline depends upon the configuration of the pipeline and how it is designed and what are the tasks that are present inside the respective jobs.

For showcasing the implementation/operation of the CI/CD pipeline with the approach discussed in the previous section, here is a small sample pipeline created in Jenkins, configured in the server which fetches the data from a central repository and does the basic testing and then deploy the code to the production server.

In Fig. 4, the pipeline is triggered automatically, using the GITSCM Polling, after the code is pushed to the central repository (GitHub in our case). Fig. 4 shows the interactive step integration just for the purpose of manual confirmation of deployment to the production. Fig. 5 and Fig. 6 respectively show the screenshots of the deploying in production and successful deployment in the production environment. Fig. 7 contains the logs accessed after the successful deployment of the Jenkins pipeline. The results exhibited in these figures show the successful implementation of the DevOps pipeline along with the required results and final deployment.

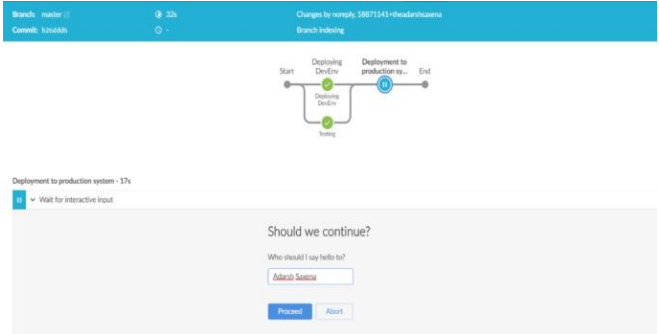

Fig. 4. CI/CD Pipeline triggered from GITSCM Polling

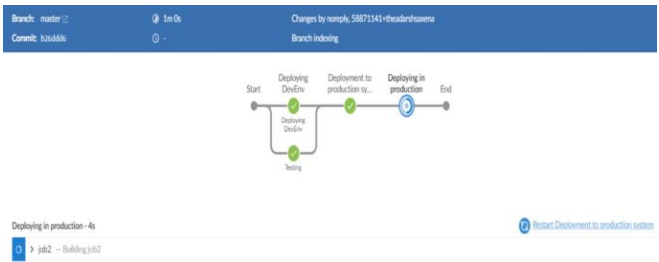

Fig. 5. Deployment in production job running in a pipeline

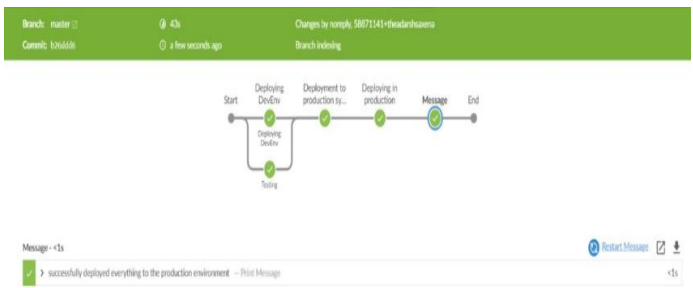

Fig. 6. Pipeline is executed successfully

```
> git checkout -f 39324ba4a2adb632552105951c511bff65139008 # timeout=10
Commit message: "Merge commit '03eb025912949638b8b020297d514f0c58a6004a' into HEAD"
> git rev-list --no-walk 502a8c8018875fb70b8247d7d7f3809b52e38ffc # timeout=10
[Job2] $ /bin/sh -xe /tmp/jenkins14696877035421926260.sh
+ sudo cp -v -r -f any.html index.html Jenkinsfile /var/lib/containers/storage/volumes/prodenv/_data
'any.html' -> '/var/lib/containers/storage/volumes/prodenv/_data/any.html'
'index.html' -> '/var/lib/containers/storage/volumes/prodenv/_data/index.html'
'Jenkinsfile' -> '/var/lib/containers/storage/volumes/prodenv/_data/Jenkinsfile'
Finished: SUCCESS
```

Fig. 7. Basic logs accessed from the log of Jenkins pipeline deployment

## V. CONCLUSION AND FUTURE WORK

A pipeline is created to demonstrate the implementation of the DevOps strategy that can be implemented in the real world. The contribution of this paper is to remove the ambiguity and conflicts in the portrayal of the DevOps by demonstrating the implementation strategy of the DevOps Pipeline in practice for better automation and faster release of the software using the Infrastructure as a Code and Pipeline as a Code. The pipeline as a code (for pipeline creation) gives much better flexibility in pipeline deployment and better collaboration between different teams. Pipeline as a code will also allow the teams to implement DevOps practices at a much deeper level within the organization.

The paper can be extended to include Docker and Kubernetes for running for the Jenkins server on top of it. Moreover, the distributed Jenkins cluster can be used by using the docker or even the AWS instances for the nodes of the Jenkins cluster. When using AWS instances for the Jenkins cluster nodes, terraform IaC can be used for the creation of the nodes, which provides a better management and replication facility. The integration of other DevOps related tools in the CI/CD pipeline would let the pipeline come up with much greater capabilities toward continuous integration and continuous delivery.